\documentclass[12pt]{article}
\usepackage{bm}
\newcommand{\bbeta}{{\mbox{\boldmath $\beta$}}}

\newcommand{\bphi}{{\mbox{\boldmath $\phi$}}}

\newcommand{\ba}{\begin{array}}
\newcommand{\p}[1]{(\ref{#1})}
\newcommand{\ea}{\end{array}}

\def\bbox{{\,\lower0.9pt\vbox{\hrule \hbox{\vrule height 0.2 cm
\hskip 0.2 cm \vrule height 0.2 cm}\hrule}\,}}
\newcommand{\dsl}{\pa \kern-0.5em /}

\newcommand{\ep}{\epsilon}

\newcommand{\nn}{\nonumber \\}

\def\Tr{{\rm Tr\,}}

\def\CM{{\cal M}}

\def\ds{\raise.15ex\hbox{/}\kern-.57em\partial}
\def\Ds{\,\raise.15ex\hbox{/}\mkern-13.5mu D}
\begin{document}

\begin{flushright}
{\tt hep-th/0602197}
\end{flushright}

\vspace{5mm}

\begin{center}
{{{\Large \bf Low Energy Dynamics of Monopoles in Supersymmetric
    Yang-Mills Theories with Hypermultiplets}}\\[12mm]
{Chanju Kim}\\[1mm]
{\it Department of Physics, Ewha Womans University,
Seoul 120-750, Korea}\\
{\tt cjkim@ewha.ac.kr}
}
\end{center}
\vspace{10mm}

\begin{abstract}
We derive the low energy dynamics of monopoles and dyons in $N=2$
supersymmetric Yang-Mills theories with hypermultiplets in arbitrary
representations by utilizing a collective coordinate expansion. We
consider the most general case that Higgs fields both in the vector
multiplet and in the hypermultiplets have nonzero vacuum expectation
values. The resulting theory is a supersymmetric quantum mechanics
which has been obtained by a nontrivial dimensional reduction of
two-dimensional (4,0) supersymmetric sigma models with potentials.
\end{abstract}

%{\it{Keywords}} :

\newpage

\section{Introduction}
In supersymmetric Yang-Mills theories (SYM) with extended supersymmetry,
there are many BPS monopole and dyon states. At weak coupling, their
low-energy dynamics can be understood semiclassically by studying the
moduli space of classical BPS monopole solutions. It turns out the dynamics is
governed by some kind of a supersymmetric quantum mechanics \cite{tongreview}.

The simplest case, where only a single adjoint Higgs field has a nonvanishing
vacuum expectation value, was analyzed in Refs.~\cite{gaunt}-\cite{gibbons}.
When a second adjoint Higgs is also nonvanishing,
there are BPS states with electric and
magnetic charge vectors that are not parallel \cite{bergman}-\cite{bak}.
In this case,
the low-energy dynamics is governed by a supersymmetric quantum mechanics
with potential terms \cite{tong}-\cite{gkpy}, which can be obtained by
a non-trivial ``Scherk-Schwarz'' dimensional reduction of
two-dimensional (4,0) supersymmetric sigma models \cite{gkly}.
This has been studied in both $N=2$ and $N=4$ theories through
direct derivation using collective coordinate approach
and/or indirect argument based on supersymmetric considerations.
In particular, in \cite{gkly}, the low-energy dynamics was derived in
$N=2$ and $N=4$ SYM with hypermultiplets when the two adjoint Higgs fields 
are nonvanishing.

One can further investigate the theory with hypermultiplets by considering
the case that the scalars in the hypermultiplets also acquire nonzero
expectation values while maintaining a nontrivial Coulomb branch.
This is possible when the hypermultiplets are massless and the representations
contain zero-weight vectors. The corresponding supersymmetric quantum
mechanics was derived when the hypermultiplets are in real
representations \cite{gkly}. In deriving this, it was crucial that there are
three complex structures on the index bundle associated with the matter
fermions in the real representation. When the representation is not
real, however, the index bundle is equipped with only a single complex
structure in general and the low-energy dynamics was not considered.
In this paper, we will address this issue and obtain the supersymmetric
quantum mechanics, which will complete the derivation of the most general
low-energy dynamics of BPS monopoles and dyons in $N=2$ and $N=4$ SYM
with hypermultiplets in arbitrary representations.

The plan of the rest of this paper is as follows. In section 2,
we briefly review the monopole dynamics in pure $N=2$ SYM to fix notations.
Section 3 is the main part of the paper.
We consider $N=2$ SYM with hypermultiplets in
arbitrary representations and derive the low-energy dynamics of monopoles
when scalars in the hypermultiplets additionally have nonzero
vacuum expectation values while maintaining a nontrivial Coulomb branch.
We conclude in section 4.

\section{Monopole Dynamics in Pure $N=2$ Super Yang-Mills Theory}
In this section, we briefly review the dynamics of monopoles in
pure $N=2$ SYM. Details can be found in \cite{gkpy,gkly}.
The Lagrangian of $N=2$ SYM is
\begin{eqnarray}\label{susylag}
L_0&=&-\Tr\Biggl\{-{1\over 4}F_{MN}F^{MN}+{1\over 2}D_M\Phi^ID^M\Phi^I
-{1\over 2}[\Phi^1,\Phi^2]^2 \nonumber \\ &&
\hskip 1cm - i\bar\chi\gamma^M D_M\chi + i\bar \chi[\Phi^1,\chi]
-\bar\chi\gamma_5[\Phi^2,\chi]\Biggr\} ,
\end{eqnarray}
where $\Phi^I$, $I=1,2$ denote the two real Higgs fields,
$D_M\Phi^I=\partial_M\Phi^I+[A_M,\Phi^I]$, $\chi$ is a Dirac spinor
and all fields are in the adjoint representation of the gauge group $G$.
The anti-hermitian generators of the Lie algebra
${\cal G}$ are normalised so that
$\Tr t^a t^b=-\delta^{ab}$. Our metric has mostly minus signature
and $\gamma_5=i\gamma_0\gamma_1\gamma_2\gamma_3$.
The classical vacuum satisfy $[\Phi^1,\Phi^2]=0$ and thus $\Phi^I$ lie
in the Cartan subalgebra of $G$: $\Phi^I={\bphi}^I\cdot {\bf H}$.
We will only consider vacua where the symmetry is maximally
broken to $U(1)^r$ where $r = \mbox{rank }G$.
For a given vacuum electric and magnetic charge two-vectors are
defined by
\begin{eqnarray}
Q^I_e &=& -\Tr\oint \hat n\cdot \vec E \,\Phi^I={\bphi}^I\cdot {\bf q}, \nn
Q^I_m &=& -\Tr\oint \hat n\cdot \vec B \,\Phi^I={\bphi}^I\cdot{\bf g},
\end{eqnarray}
where we have introduced the electric and magnetic charge vectors,
\begin{eqnarray}
{\bf q}&=&n^m_e {\bbeta^m} ,\nn
{\bf g}&=&4\pi n^m_m{\bbeta_m^*} ,
\end{eqnarray}
respectively. ${\bbeta^m}$ are the simple roots and
${\bbeta_m^*}$ are the simple co-roots of $\cal G$, and
$n^m_m$ are the topological winding numbers and $n^m_e$
are, in the quantum theory, the electric quantum numbers.

There is a classical mass bound given by \cite{fh,leeyi}
\begin{eqnarray}\label{bpse}
M&\ge& \max |Z_\pm| \nn
&\equiv&\max|(Q^1_e - Q^2_m) \pm i(Q^1_m + Q^2_e)|.
\end{eqnarray}
Only $Z_-$ appears as a central charge in the $N=2$ supersymmetry algebra
and half-BPS states satisfy $M=|Z_-|$ \cite{wittenolive,gkpy}.
Thus BPS solitons can only have charges satisfying $|Z_-|\ge |Z_+|$.
This bound is saturated when
\begin{eqnarray}\label{bpsbound}
\vec E&=&\pm \vec Da ,\nn
\vec B&=&\vec Db ,
\end{eqnarray}
where we have defined the rotated Higgs fields via
\begin{eqnarray}\label{aandb}
a&=&\cos\alpha\Phi^1-\sin\alpha\Phi^2 ,\nn
b&=&\sin\alpha \Phi^1+\cos\alpha\Phi^2 ,
\end{eqnarray}
and the angle $\alpha$ is constrained to be
\begin{equation}\label{anglevalue}
\tan\alpha = {Q_m^1 \mp Q_e^2\over Q_m^2 \pm Q_e^1}\, .
\end{equation}
The second equation in (\ref{bpsbound}) is the usual
BPS equation for a single Higgs field of which the solutions are
usual BPS monopoles. For a given solution of the the second equation,
the first equation has a unique solution for specified asymptotic
behavior of $a$. The solutions to the general equations can thus be viewed
as electrically dressed solutions to the BPS monopoles.

In terms of the vectors ${\bf a},{\bf b}$, which are defined through
(\ref{aandb}), the mass bound is given by
\begin{eqnarray}\label{boundalt}
M\ge \max (\pm{\bf a}\cdot {\bf q} +{\bf b}\cdot{\bf g} ),
\end{eqnarray}
which can be obtained by noting that (\ref{anglevalue}) can be recast as
\begin{equation}\label{constraintv}
{\bf b}\cdot {\bf q}= \pm{\bf a}\cdot{\bf g}.
\end{equation}

In deriving the low-energy dynamics, we treat these dyons as particular
excited states of the monopole dynamics. We thus begin with a given
magnetic charge vector ${\bf g}$ and fixed Higgs expectation values $\Phi^I$.
Setting ${\bf q}=0$ then fixes the angle $\alpha$ and the fields $a,b$ defined
in \p{aandb}. From (\ref{constraintv}), it also means that ${\bf a}$ is
orthogonal to the magnetic charge,
\begin{equation}\label{adotg}
{\bf a\cdot g}=0.
\end{equation}
The collective coordinate expansion then begins with a static purely magnetic
solution to the equation $B_i=D_ib$.
The dynamical effect of the second Higgs field is treated as
a perturbation of this solution.
The collective coordinate expansion can be considered to be an
expansion in $n=n_\partial + \frac12 n_f$, where $n_\partial$ is the number
of time derivatives and $n_f$ is the number of fermions.
The equations of motion of the low-energy effective action
will be of order $n=2$, so we will solve
the equations of motion of the field theory to order
$n=0$, $n=\frac12$ and $n=1$. To incorporate the affects of the second Higgs
field we will also assume that $a$ is of order $n=1$.

Since the collective coordinate expansion is constructed about
solutions of the ordinary BPS equation for a single Higgs field
$B_i=D_i b$, we summarize some aspects
of the geometry of the moduli spaces of solutions following
\cite{harvstrom,gaunt}.

We first define a connection $W_\mu$ on $R^4$ that is
translationally invariant in the four direction via
$W_\mu=(A_i,b)$ and field strength $G_{\mu\nu} = [D_\mu, D_\nu]$ with
$D_\mu = \partial_\mu + [W_\mu,]$.
then the BPS equations can be recast as self-duality
equations for $W_\mu$,
\begin{equation}
G_{\mu\nu}={1\over 2}\epsilon_{\mu\nu\rho\sigma}G_{\rho\sigma}.
\end{equation}

Denote the moduli space of solutions to the BPS
equations within a given topological class $k$ by ${\cal M}_k$.
A natural set of coordinates is provided by the moduli $z^m$ that specify
the most general gauge equivalence class of solutions $W_\mu(x,z)$.
The zero modes $\delta_m W_{\mu}$
about a given solution satisfy the linearized BPS equation
\begin{equation}
D_{[\mu}\delta_m W_{\nu]}=
{1\over 2}\epsilon_{\mu\nu\rho\sigma}D_{\rho}\delta_m W_{\sigma},
\end{equation}
as well as
\begin{equation}\label{orth}
D_\mu\delta_m W_\mu=0.
\end{equation}
A natural metric on ${\cal M}_k$ is
\begin{equation}\label{metrict}
g_{mn}=-\int d^3x \Tr(\delta_m W_\mu\delta_n W_\mu).
\end{equation}
Then (\ref{orth}) implies that the zero mode is orthogonal to gauge modes.

The zero modes are in general written as
\begin{equation}
\delta_m W_\mu=\partial_m W_\mu-D_\mu \eta_m,
\end{equation}
where the gauge parameters $\eta_m(x,z)$ are chosen to
satisfy (\ref{orth}). Then, on $\CM_k$, $\eta_m$
define a natural connection with covariant derivative
\begin{equation}
s_m=\partial_m+[\eta_m,\ ],
\end{equation}
and field strength
\begin{equation}
\phi_{mn}=[s_m,s_n].
\end{equation}
The pair $(W_\mu(x,z),\eta_m(x,z))$ defines a natural connection on
$R^4\times\CM_k$. The components of the field strength are given by
$G_{\mu\nu}$, $\phi_{mn}$ and the mixed components are given by
\begin{equation}\label{cross}
[s_m,D_\mu]=\delta_m W_\mu.
\end{equation}
They satisfy the following identities:
\begin{eqnarray}
s_mG_{\mu\nu}&=&2D_{[\mu}\delta_mW_{\nu]},\nn
D_\mu\phi_{mn}&=&-2s_{[a}\delta_{b]}W_\mu,\nn
\phi_{mn}&=&2(D_\mu D_\mu)^{-1}[\delta_m W_\nu,\delta_n W_\nu].
\end{eqnarray}

The Christoffel connection associated with the metric
(\ref{metrict}) can be written in the form:
\begin{equation}
\Gamma_{mnk}=g_{ml}{\Gamma^l}_{nk}=-\int d^3x
\Tr(\delta_m W_\mu s_k\delta_n W_\mu).
\end{equation}
The hyper-K\"ahler structure on $R^4$ gives rise to a hyper-K\"ahler
structure on ${\cal M}_k$. The three complex structures can
be written
\begin{equation}\label{compstr}
J^{(s)n}_m=-g^{np}\int d^3x {J^{(s)}}_{\mu\nu} \Tr
(\delta_m W_\mu\delta_p W_\nu),
\end{equation}
which implies
\begin{equation}\label{hkid}
J^{(s)n}_m\delta_n W_\mu=-J^{(s)}_{\mu\nu}\delta_m W_\nu.
\end{equation}

Now we turn to the zero modes of the adjoint fermions.
On the Euclidean space $R^4$, we introduce Hermitian gamma matrices,
\begin{equation}
\Gamma_i=\gamma_0\gamma_i,\qquad\Gamma_4=\gamma_0,
\end{equation}
satisfying $\{\Gamma_\mu,\Gamma_\nu\}=2\delta_{\mu\nu}$ and
define $\Gamma_5=\Gamma_1\Gamma_2\Gamma_3\Gamma_4$. The fermion
zero modes are time independent solutions of the Dirac equation
in the presence of a BPS monopole,
\begin{equation}
\Gamma_\mu D_\mu\chi=0.
\end{equation}
They are necessarily anti-chiral. The monopole breaks 1/2 of the supersymmetry
and the unbroken supersymmetry can be used to pair the bosonic
and fermionic zero modes via
\begin{equation}\label{fizz}
\chi_m = \delta_m W_\mu\Gamma^\mu\epsilon_+,
\end{equation}
where $\epsilon_+$ is a c-number chiral spinor that can be chosen to satisfy
\begin{equation}\label{sunday}
\epsilon_+^\dagger\epsilon_+=1, \qquad
J^{(3)}_{\mu\nu}=-i\epsilon_+^\dagger\Gamma_{\mu\nu}\epsilon_+ .
\end{equation}
Using (\ref{hkid}) we deduce that the fermionic zero modes satisfy
\begin{equation}\label{fzmid}
J_m^{(3)n}\chi_n=i\chi_m ,
\end{equation}
and hence that two bosonic zero modes
are paired with one fermionic zero mode \cite{callias}.

For later use, we discuss more on the complex structures. The charge
conjugation of the spinor $\chi$ is defined as
\begin{equation}
\chi^c\equiv C\bar\chi^T =C(\gamma^0)^T\chi^*
\end{equation}
where the charge-conjugation matrix $C$ satisfies,
\begin{equation}
CC^*=-1, \qquad C\Gamma_M^T=-\Gamma_MC .
\end{equation}
Then with the c-number spinor $\epsilon'_+ \equiv C\epsilon_+^*$, we see that
$\delta_m W_\mu \Gamma_\mu \epsilon'_+$ are also zero modes and
can be expressed as a linear combination of original zero modes since
$\chi$ (and $W$) is in a real representation of the gauge group, i.e.,
\begin{equation} \label{cmn}
\delta_m W_\mu \Gamma_\mu \epsilon'_+ =
{\cal C}_m^{\;\;\;k}\delta_k W_\mu \Gamma_\mu \epsilon_+ ,
\end{equation}
where the matrix ${\cal C}$ can be chosen to be
anti-symmetric and unitary so that ${\cal C}^2=-1$.
By taking the complex conjugate of (\ref{fzmid}),
it follows that ${\cal C}$ anticommutes with $J^{(3)}$,
\begin{equation}
{\cal C}J^{(3)}=-J^{(3)}{\cal C} .
\end{equation}
This matrix $\cal C$ generates a second complex structure on the
moduli space which we also denote by $J^{(2)}$.
Defining $J^{(1)}$=$J^{(2)}J^{(3)}$ we obtain the hyper-K\"ahler
structure of the monopole moduli space
which can be taken to be the same as \p{compstr} by an appropriate choice
of complex structures on $R^4$.

With the above formalism on moduli space, it is now quite a simple matter
to derive the low-energy effective action of pure $N=2$ SYM. First, we
rewrite the Lagrangian in terms of $b,a$ rather than $\Phi^1,\Phi^2$,
\begin{eqnarray}\label{susylagab}
L&=&-\Tr\Biggl\{-{1\over 4}F_{MN}F^{MN}+{1\over 2}D_M aD^M a +
{1\over 2}D_M bD^M b
-{1\over 2}[a,b]^2 \nonumber \\ &&
\hskip 1cm - i\bar\chi\gamma^M D_M\chi +i\bar \chi[b,\chi]
+\bar\chi\gamma_5[a,\chi]\Biggr\} ,
\end{eqnarray}
where $\chi$ has now been redefined as the field rotated by the
angle $(\alpha-\pi/2)/2$. Then the following ansatz solve the equations
of motion to order $n=1$ \cite{gkly}:
\begin{eqnarray}\label{ansatztwo}
W_\mu&=&W_\mu(x,z(t)),\nn
\chi&=&\delta_mW_\mu\Gamma^\mu\epsilon_+\tilde\lambda^m(t),\nn
A_0&=&\dot z^m\eta_m-i\phi_{mn}\tilde\lambda^{\dagger m}\tilde\lambda^n,\nn
a&=&\bar a +i\phi_{mn}\tilde\lambda^{\dagger m}\tilde\lambda^n,
\end{eqnarray}
where
\begin{equation}\label{cat}
D_\mu \bar a = -G^m\delta_m W_\mu ,
\end{equation}
and $G^m$ is a linear combination of the $r$
tri-holomorphic Killing vector fields ${\bf K}$ on $\CM_k$ corresponding
to the $U(1)^r$ gauge transformations
\begin{equation}
G={\bf a}\cdot {\bf K}.
\end{equation}

Because of (\ref{fzmid})
the complex fermionic Grassmann odd collective coordinates
$\tilde\lambda^m$ are not independent and satisfy
\begin{equation}\label{fccid}
-i\tilde\lambda^m J_m^{(3)n}=\tilde\lambda^n .
\end{equation}
Real independent $\lambda^m$ can be defined via
\begin{equation} \label{reallambda}
\lambda^m=\sqrt{2}\left(\tilde\lambda^m+(\tilde\lambda^m)^\dagger\right) .
\end{equation}

After substituting the ansatz into the action, one finds that
the low-energy effective action becomes \cite{gkly}
\begin{equation}\label{actiontoo}
S=\frac12\int dt[\dot x^m \dot x^n g_{mn}+
ig_{mn} \lambda^m D_t \lambda^n - G^m G^n g_{mn}
- iD_m G_n  \lambda^m \lambda^n]
-{\bf b}\cdot {\bf g} ,
\end{equation}
which was first given in \cite{gkpy} based on supersymmetry considerations.

\section{Inclusion of Hypermultiplets}

We now consider the low-energy dynamics of monopoles in
$N=2$ SYM with a hypermultiplet in an arbitrary representation.
This was first studied in \cite{sethi,cederwall,gauntharv} in the case that
only a single adjoint Higgs field has a non-trivial expectation value.
It was then generalized in \cite{gkly} to the case that
both of the adjoint Higgs in the vector multiplet
have non-vanishing expectation values. In \cite{gkly}, the low-energy
dynamics was also derived when additional scalar vevs in
the hypermultiplet are turned on. In deriving this, it was necessary
to assume that the hypermultiplet is in a real representation to utilize
complex structures of the index bundle associated with the matter fermions.
Here, we derive the low-energy effective theory for the most general
case, namely when the hypermultiplet is in an arbitrary representation and
nontrivial scalar vevs of the hypermultiplet are turned on.

The massless hypermultiplet contribution to the Lagrangian is given by
\begin{eqnarray}\label{matterlag}
L_H&=&\frac12 D_K M^{\dagger } D^K  M
      + i \bar\Psi\gamma^K D_K \Psi
      - \bar\Psi(-i\Phi_1 - \gamma_5 \Phi_2)\Psi \nn
   &&+ M^{\dagger 1}\bar\chi\Psi + \bar\Psi\chi M_1
     + iM^{\dagger 2}\bar\chi^c\gamma_5\Psi
     + i\bar\Psi\gamma_5\chi^c M_2 \nn
   &&+\frac12 M^{\dagger} (\Phi_1^2 + \Phi_2^2) M
     +\frac18 (M^{\dagger } t^\alpha \tau_s M)^2,
\end{eqnarray}
where $M$ is a doublet of complex scalars $(M_1,M_2)^T$,
$t^\alpha$ are anti-hermitian generators in the matter
representation, $\tau_s$ are Pauli matrices, and
$\chi^c$ is the charge conjugation of $\chi$. Since we will assume
that $M_i$'s have nonzero vevs, the hypermultiplet is necessarily massless
and its representation should contain a zero-weight vector so that the
$U(1)$ gauge symmetries of the Coulomb phase are left intact by turning
on the vevs.

Before discussing the low-energy dynamics of the system, we briefly
summarize some aspects of the geometry of the index bundle
defined by the fermion zero modes.
The zero modes of matter fermion $\Psi$ satisfy the the Dirac equation in the
background of a monopole configuration
\begin{equation}\label{diraceq}
\Gamma_\mu D_\mu\gamma_5\Psi =0    ,
\end{equation}
and are chiral. Let $\Psi_A(x,z)$, $A=1\dots l$ be a basis of the
fermion zero modes in monopole background specified by the moduli $z$
satisfying
\begin{equation}\label{dog}
\int d^3 x \Psi^{\dagger}_{\bar A}\Psi_B
\equiv \langle\Psi_{\bar A}|\Psi_B\rangle=
\delta_{\bar AB}  ,
\end{equation}
where $\Psi^\dagger_{\bar A}\equiv(\Psi_A)^\dagger$.
Note that the following completeness relation holds:
\begin{equation}\label{completeness}
|\Psi_A\rangle\delta^{A \bar B}\langle\Psi_{\bar B}|
  +\Pi +{1-\Gamma_5\over 2}=1 ,
\end{equation}
where the operator $\Pi$ projects onto the chiral non-zero modes and
has the form
\begin{equation}
\Pi=\gamma_5{\Ds}{1\over D_\mu D_\mu}{\Ds}\gamma_5{1+\Gamma_5\over 2} .
\end{equation}
The fermion zero modes define an index bundle with a connection
\begin{equation}\label{mattercon}
{A_m}_{\bar AB}=\langle\Psi_{\bar A}|s_m\Psi_B\rangle ,
\end{equation}
and the corresponding field strength is written in the form
\begin{equation}
F_{mn\bar AB}=\langle s_m\Psi_{\bar A}|\Pi s_n\Psi_B\rangle
-\langle s_n\Psi_{\bar A}|\Pi s_m\Psi_B\rangle
+\langle \Psi_{\bar A}|\phi_{mn}\Psi_B\rangle  .
\end{equation}
Since the connection one-form is unitary,
the structure group of the index bundle is generically $U(l)$.
The index bundle thus admits a covariantly constant complex
structure $I$ with K\"ahler form
\begin{equation} \label{cI}
I_{A\bar B}=i\delta_{A \bar B}.
\end{equation}

Now the collective coordinate expansion can be done. After a suitable
chiral rotation as in the previous section, the ansatz solving the equations
of motion to order $n=1$ is \cite{gkly}
\begin{eqnarray}\label{ansatzthree}
W_\mu&=&W_\mu(x,z(t)),\nn
\chi&=&\delta_mW_\mu\Gamma^\mu\epsilon_+\tilde\lambda^m(t),\nn
A_0&=&\dot z^m\eta_m-i\phi_{mn}\tilde\lambda^{\dagger m}\tilde\lambda^n
+\frac{i}{D^2}(\Psi^\dagger t^\alpha\Psi t^\alpha),\nn
a&=&\bar a +i\phi_{mn}\tilde\lambda^{\dagger m}\tilde\lambda^n
+\frac{i}{D^2}(\Psi^\dagger t^\alpha\Psi t^\alpha), \nn
\Psi&=&\psi^A(t)\Psi_A ,\nn
M_1&=&\bar M_1-\frac{2}{D^2}(\bar\chi\Psi) ,\nn
M_2&=&\bar M_1-\frac{2i}{D^2}(\bar\chi^c\gamma_5\Psi) ,
\end{eqnarray}
where $\psi^A(t)$ is the Grassmann odd complex collective coordinates
for the matter fermion zero modes.
$\bar a$ satisfies (\ref{cat}) and $\bar M_{1,2}$ are order $n=1$ and solve
\begin{equation}\label{fact}
D^2 \bar M_{1,2}=0 .
\end{equation}
After substituting this ansatz into the field theory action,
the $\bar M$-independent terms give rise to the supersymmetric
quantum mechanics derived in \cite{gkly}:
\begin{eqnarray}
{\cal L}_1&=&{1\over 2} \biggl( g_{mn} \dot{z}^m \dot{ z}^n +
ig_{mn} \lambda^m D_t \lambda^n
 - g^{mn} G_m G_n - iD_m G_n  \lambda^m \lambda^n \nn
&&+i\psi^a{\cal D}_t\psi^a + {1\over 2}F_{mn ab}\lambda^m
\lambda^n\psi^a\psi^b - i T_{ab}\psi^a\psi^b\biggr) -{\bf b\cdot g} .
\label{action}
\end{eqnarray}
where we traded off complex $\psi^A$'s in favor of real $\psi^a$'s as in
(\ref{reallambda}) and
\begin{equation}
{\cal D}_t\psi^a=\dot \psi^a +{{A_m}^a}_b\dot z^m\psi^b .
\end{equation}
$T$ is defined by
\begin{equation}\label{tee}
T_{\bar AB}= \langle\Psi_{\bar A}|\bar a\Psi_B\rangle ,
\end{equation}
and is anti-Hermitian in the real basis, $T_{ab} = -T_{ba}$.
Furthermore, it satisfies
\cite{gkly}
\begin{equation}\label{tues}
T_{\bar AB;m}= F_{mn \bar AB}G^n .
\end{equation}

In the following we derive $\bar M$-dependent terms which are the main
result of this paper.

\subsection{Bosonic potential}
First we note that, given (\ref{fact}), $\Ds \bar M_i \epsilon_+$ and
$\Ds \bar M_i \epsilon'_+$ are fermion zero modes and hence can be expanded
in terms of the basis $\Psi_A$:
\begin{eqnarray} \label{dhem}
\Ds \bar M_i \epsilon_+ &=&-\gamma_5\sqrt{2}K_i^A(z)\Psi_A ,\nn
\Ds \bar M_i \epsilon'_+ &=& -\gamma_5\sqrt{2}K_i^{\prime A}(z)\Psi_A .
\end{eqnarray}
The quantities $K_i^A$ and $K_i^{\prime A}$ define sections
on the dual of the index bundle over the monopole moduli space.
Then $K_i^A$ and $K_i^{\prime A}$ are orthogonal to each other,
\begin{equation} \label{kkij}
K_i^{*A} K_j^{\prime A} = 0.
\end{equation}
To see this, write
\begin{equation} \label{kikj}
2 K_i^{*A} K_j^{\prime A}
= \int d^3x\, (\Ds \bar M_i \epsilon_+)^\dagger (\Ds \bar M_j \epsilon'_+),
\end{equation}
where we used the orthogonality of the zero modes $\Psi_A$.
In the right hand side of the equation, we have the expression
\begin{equation}
\epsilon_+^\dagger \Gamma_\mu \Gamma_\nu \epsilon'_+
= \epsilon_+^\dagger (\delta_{\mu\nu} + \Gamma_{\mu\nu}) \epsilon'_+.
\end{equation}
Since the c-number chiral spinor $\epsilon_+$ is orthogonal to its
charge-conjugated one $\epsilon'_+$, which can easily be verified by using
(\ref{sunday}), only the antisymmetric part survives and hence (\ref{kikj})
becomes
\begin{equation}
2 K_i^{*A} K_j^{\prime A}
= \frac12 \epsilon_+^\dagger \Gamma_{\mu\nu} \epsilon'_+
   \int d^3x\, D_{[\mu} \bar M_i^\dagger D_{\nu]} \bar M_j .
\end{equation}
After an integration by parts, this can be written as a sum of a
vanishing boundary integral and the term containing
the field strength $G_{\mu\nu} = [D_\mu,D_\nu]$. But this is self-dual
and goes to zero when multiplied by
$\epsilon_+^\dagger \Gamma_{\mu\nu} \epsilon'_+$ since $\epsilon_+$ is chiral.
This establishes (\ref{kkij}). Furthermore, it is clear from the definition
that $K_1$ and $K_2$ have the same magnitude as $K'_1$ and $K'_2$, i.e.,
\begin{equation} \label{kkij2}
K_1^{*A} K_1^A = K_1^{\prime *A} K_1^{\prime A}, \qquad
K_2^{*A} K_2^A = K_2^{\prime *A} K_2^{\prime A}.
\end{equation}

Now we are ready to deal with the $\bar M$-dependent bosonic potential terms,
which arise from the kinetic terms of $M$ in (\ref{matterlag}).
Using the similar line of argument as above, we find that they reduce to
\begin{eqnarray} \label{bopo}
{\cal L}_{Hb} &\equiv&
  \frac12 \int d^3x\;D_\mu  \bar M_i^\dagger D_\mu  \bar M_i \nonumber\\
  &=& \frac12 \int d^3x\;(\Ds \bar M_i \epsilon_+)^\dagger
       \Ds \bar M_i \epsilon_+ \nonumber \\
  &=& K_i^{*A}K_i^A.
\end{eqnarray}
(The cross terms which are linear in $\bar M$ vanish due to (\ref{fact}).)

It turns out to be convenient to define
\begin{equation} \label{va}
v^A = K_1^A - K_2^{\prime A}.
\end{equation}
Then using (\ref{kkij}) and (\ref{kkij2}), we can rewrite (\ref{bopo}) as
\begin{equation} \label{lhb}
{\cal L}_{Hb} = v^{*A}v^A = \frac12 v^a v_a,
\end{equation}
where, as before, we rewrote the complex quantities $v^A$ in terms of real
quantities $v^a$ by expanding
\begin{equation} \label{realv}
v^A=\frac{1}{\sqrt{2}}\left( v^{2A-1}+iv^{2A}\right) .
\end{equation}

\subsection{Fermion bilinear terms}

Fermion bilinear terms are
\begin{equation} \label{lfb}
{\cal L}_{Hf} \equiv \int M^{\dagger 1}\bar\chi\Psi + \bar\Psi\chi M_1
             + iM^{\dagger 2}\bar\chi^c\gamma_5\Psi
             + i\bar\Psi\gamma_5\chi^c M_2 .
\end{equation}
With the relation $\delta_m W_\mu = [s_m, D_\mu]$, we find
\begin{equation} \label{chih}
\chi \bar M_1 = \tilde\lambda^m s_m \Ds \bar M_1 \epsilon_+   +\cdots,
\end{equation}
where the ellipsis denote terms of the form $\Ds(\ldots)$ which do not
contribute any new terms in the low energy dynamics since $\Psi$ in the
Lagrangian is chiral and satisfies the Dirac equation. Using the relation
(\ref{cmn}), this can also be written as
\begin{equation} \label{chih2}
\chi \bar M_1 = -\tilde\lambda^m {\cal C}_m^{\;\;\;n} s_n
             \Ds \bar M_1 \epsilon'_+ +\cdots .
\end{equation}
The existence of two alternative expressions for the same quantity is
basically related to the hyper-K\"ahler structure of the moduli space, as
mentioned in section 2. We will see shortly that this plays crucial
roles in constraining the form of the effective Lagrangian so that
it becomes supersymmetric.
Using (\ref{dhem}), we see that the term containing $\bar M_1$ in
(\ref{lfb}) becomes
\begin{eqnarray} \label{fb1}
\int\bar\Psi\chi \bar M_1
  &=& i\sqrt2 \tilde\lambda^m \psi^{\bar A} \nabla_m K_{1 \bar A} \nn
  &=& -i\sqrt2 \tilde\lambda^m
          {\cal C}_m^{\;\;\;n} \psi^{\bar A} \nabla_n K'_{1 \bar A}.
\end{eqnarray}
Similarly, for $-i\gamma_5\chi^c \bar M$,
\begin{eqnarray} \label{chich}
-i\gamma_5\chi^c \bar M_2 &=& \tilde\lambda^{\dagger m} s_m \Ds M_2 \epsilon'_+
                      + \dots \nn
                   &=& \tilde\lambda^{\dagger m} {\cal C}_m^{\;\;\;n} s_n
                      \Ds M_2 \epsilon_+ + \dots ,
\end{eqnarray}
which gives
\begin{eqnarray} \label{fb2}
\int i\bar\Psi\gamma_5\chi^c \bar M_2
&=& -i\sqrt2 \tilde\lambda^{\dagger m} \psi^{\bar A} \nabla_m K'_{2 \bar A}\nn
&=& -i\sqrt2 \tilde\lambda^{\dagger m} \psi^{\bar A} {\cal C}_m^{\;\;\;n}
\nabla_n K_{2 \bar A}.
\end{eqnarray}

Taking into account the complex conjugates of (\ref{fb1}) and (\ref{fb2}),
we find that the fermion bilinear terms become
\begin{equation} \label{sf1}
{\cal L}_{Hf} = i\sqrt2 ( \tilde\lambda^m \psi^{\bar A} \nabla_m K_{1 \bar A}
              + \tilde\lambda^{\dagger m} \psi^A \nabla_m K_{1A}
              - \tilde\lambda^m \psi^A \nabla_m K'_{2A}
       - \tilde\lambda^{\dagger m} \psi^{\bar A} \nabla_m K'_{2 \bar A} ),
\end{equation}
where we used the expression without ${\cal  C}_m^{\;\;\;n}$.
In terms of the real quantities $v^a$ defined in (\ref{va}) and $\lambda^m$,
it can be reshuffled to
\begin{eqnarray} \label{sf2}
{\cal L}_{Hf} &=& i \lambda^m \psi^a \nabla_m v_a \\
    && -i\sqrt2 ( \tilde\lambda^m \psi^A \nabla_m K_{1A}
              - \tilde\lambda^m \psi^{\bar A} \nabla_m K'_{2 \bar A}
              - \tilde\lambda^{\dagger m} \psi^A \nabla_m K'_{2A}
       + \tilde\lambda^{\dagger m} \psi^{\bar A} \nabla_m K_{1 \bar A} ).
\nonumber
\end{eqnarray}

Now we are going to show that each of the last four terms in the above
equation is actually zero.
First, note that (\ref{fb1}) gives a nontrivial relation
\begin{equation} \label{frel1}
(1-iJ^{(3)})\nabla \psi^{\bar A}K_{1\bar A}
= - (1-iJ^{(3)}) {\cal C}\nabla \psi^{\bar A}K'_{1\bar A}.
\end{equation}
where we used (\ref{fccid}).

Note the operators $(1\mp i J^{(3)})\nabla$ appearing in (\ref{frel1})
are holomorphic and
anti-holomorphic covariant derivatives with respect to the third complex
structure $J^{(3)}$ on ${\cal M}_k$. The reason that $J^{(3)}$ appears
in this equation is because we used the c-number
spinors $\epsilon_+, \epsilon'_+$ associated $J^{(3)}$
in considering the fermion zero modes.
There are, however, three complex structures on ${\cal M}_k$ and
hence we can obtain similar relations if we use the c-number spinors
associated with the other complex structures $J^{(2)} = {\cal C}$ and
$J^{(1)} = J^{(2)} J^{(3)}$. The corresponding c-number spinors
$\epsilon_+^{(s)}$, $s=1,2,3$ are defined by
\begin{equation} \label{eps}
\epsilon_+^{(s)\dagger}\epsilon_+^{(s)}=1, \qquad
J^{(s)}_{\mu\nu}=-i\epsilon_+^{(s)\dagger}\Gamma_{\mu\nu}\epsilon_+^{(s)},
\qquad
J^{(s)}_{\mu\nu} \Gamma_\nu \epsilon_+^{(s)} = i \Gamma_\mu \epsilon_+^{(s)} ,
\end{equation}
which generalize the relation (\ref{sunday}).
The corresponding zero modes are denoted as
\begin{equation}
\chi_m^{(s)} = \delta_m W_\mu\Gamma^\mu\epsilon_+^{(s)} .
\end{equation}
After some calculation, the explicit form of $\epsilon_+^{(s)}$ can be
obtained by using (\ref{eps}):
\begin{eqnarray} \label{eps12}
\epsilon_+^{(1)} &=& -\frac{e^{i\pi/4}}{\sqrt{2}}(\epsilon_+ + \epsilon'_+),
\nn
\epsilon_+^{(2)} &=& -\frac{e^{-i\pi/4}}{\sqrt{2}}(\epsilon_+ - i\epsilon'_+),
\end{eqnarray}
where phases of the spinors are carefully chosen so that the cyclicity
for the label of the complex structures are manifest in various relations
shown below. (We will continue omit the superscript label for quantities
associated with $J^{(3)}$.)
We also define $\epsilon_+^{\prime (s)} =C\epsilon_+^{(s)*}$.

Now let us consider the expansion
\begin{eqnarray} \label{dhems}
\Ds \bar M_i \epsilon_+^{(2)} &=&-\gamma_5\sqrt{2}K_i^{(2)A}(z)\Psi_A ,\nn
\Ds \bar M_i \epsilon_+^{\prime (2)}
&=& -\gamma_5\sqrt{2}K_i^{(2)\prime A}(z)\Psi_A .
\end{eqnarray}
From the relation (\ref{eps12}), we can express
$K_i^{(2)A}$ and $K_i^{(2)\prime A}$ as
\begin{eqnarray} \label{ki2}
K_i^{(1)A} &=& -\frac{e^{i\pi/4}}{\sqrt2} ( K_i +  K'_i)^A, \qquad
K_i^{(1)\prime A} = -\frac{e^{-i\pi/4}}{\sqrt2} ( K'_i -  K_i)^A, \nn
K_i^{(2)A} &=& -\frac{e^{-i\pi/4}}{\sqrt2} ( K_i - i K'_i)^A, \qquad
K_i^{(2)\prime A} = -\frac{e^{i\pi/4}}{\sqrt2} ( K'_i - i K_i)^A.
\end{eqnarray}
With the expansion (\ref{dhems}) for $s=2$, the condition corresponding to
(\ref{frel1}) now takes the form
\begin{equation} \label{frel2}
(1-iJ^{(2)})\nabla \psi^{\bar A}K_{1\bar A}^{(2)}
= - (1-iJ^{(2)}) J^{(1)} \nabla \psi^{\bar A}K_{1\bar A}^{(2)\prime},
\end{equation}
which, on using (\ref{frel1}) and (\ref{ki2}), can be simplified to
\begin{equation} \label{rel1}
\nabla \psi^{\bar A}K_{1\bar A}
=-i J^{(1)} \nabla \psi^{\bar A}K'_{1\bar A}.
\end{equation}
Similarly, for $s=1$ we obtain the relation
\begin{equation} \label{rel2}
(1+ iJ^{(3)}) \nabla \psi^{\bar A}K_{1\bar A}
=-(1+ iJ^{(3)}) J^{(2)} \nabla \psi^{\bar A}K'_{1\bar A}.
\end{equation}
Combining (\ref{rel1}) and (\ref{rel2}), we actually find that each
side of (\ref{rel2}) is zero separately. In other words,
\begin{equation} \label{rel3}
\tilde \lambda^{\dagger m} \nabla_m \psi^{\bar A}K_{1\bar A} =0,\qquad
\tilde \lambda^m \nabla_m \psi^{\bar A}K'_{1\bar A} =0.
\end{equation}
Finally, exactly the same kind of analysis with (\ref{chich}) shows that
\begin{equation} \label{rel4}
\tilde \lambda^{\dagger m} \nabla_m \psi^A K'_{2\bar A} =0,\qquad
\tilde \lambda^{\dagger m} \nabla_m \psi^{\bar A}K_{2\bar A} =0.
\end{equation}
(\ref{rel3}) and (\ref{rel4}) complete the proof that the second line
of (\ref{sf2}) is zero, i.e, the effective action coming from the fermion
bilinear terms become
\begin{equation} \label{lhf}
{\cal L}_{Hf} = i\lambda^m \psi^a \nabla_m v_a.
\end{equation}
From (\ref{lhb}) and (\ref{lhf}), the contribution to the effective
Lagrangian from nonvanishing hypermultiplet vevs is
\begin{equation} \label{l2}
{\cal L}_2 = \frac12 v^a v_a + i\lambda^m \psi^a \nabla_m v_a.
\end{equation}

\subsection{Low-energy effective theory}
Here we summarize the general low-energy dynamics of monopoles in $N=2$ SYM
with hypermultiplets. Collecting the terms (\ref{action}) and (\ref{l2}),
we find that the Lagrangian is given by
\begin{eqnarray} \label{fulllag}
{\cal L}&=&{1\over 2} \left( g_{mn} \dot{z}^m \dot{ z}^n +
ig_{mn} \lambda^m D_t \lambda^n
 - g^{mn} G_m G_n - iD_m G_n  \lambda^m \lambda^n \right. \nn
&&+i\psi^a{\cal D}_t\psi^a + {1\over 2}F_{mn ab}\lambda^m
\lambda^n\psi^a\psi^b - i T_{ab}\psi^a\psi^b \nn
&& \left. +v^a v_a + 2i\lambda^m \psi^a \nabla_m v_a \right) -{\bf b\cdot g}.
\end{eqnarray}
This Lagrangian has the same form as that obtained by a non-trivial
Scherk-Schwarz dimensional reduction on a two dimensional (4,0)
supersymmetric sigma model with potential \cite{hpt}. It is
invariant under $N=4$ supersymmetry transformation given by
\begin{eqnarray}
\delta z^m &=& -i\ep\lambda^m +i\ep_s {J^{(s)m}}_n \lambda^n ,\nn
\delta \lambda^m&=&(\dot z^m -G^m)\ep +{J^{(s)m}}_n(\dot z^n -
G^n)\ep_s
-i\ep_s \lambda^k \lambda^n {J^{(s)l}}_k \Gamma^m_{ln} ,\nn
\delta\psi^a&=&-{{A_m}^a}_b\delta z^m\psi^b + \ep v^a + \ep_s t_{(s)}^a ,
\end{eqnarray}
where $\epsilon, \epsilon_s$, $s=1,2,3$ are Grassmann odd parameters,
provided that the sections $t_{(s)}^a$, $s=1,2,3$ can be found
satisfying \cite{gkly}:
\begin{eqnarray} \label{convs}
J^{(s)n}_m \nabla_n v^a&=&-\nabla_m t^a_{(s)} ,\nn
(v_a t^a_{(s)})_{;m}&=&0 ,\nn
G^n v^a_{;n}&=& T_{ab} v^b,\nn
G^n t^a_{(s);n}&=& T_{ab} t^b_{(s)} .
\end{eqnarray}
In addition, the following constraints should be met: The first is
the well-known requirements that the moduli space is hyper-K\"ahler
and the curvature $F$ is of (1,1) type with respect to all three
complex structures of the manifold. Also $G$ must be a tri-holomorphic
Killing vector field, and the two form on the bundle $T$ must satisfy
\begin{equation} \label{fff}
T_{ab;m}= F_{mn ab}G^n .
\end{equation}
(\ref{fff}) is already discussed in (\ref{tues}) and,
in the following, we will show that the relations in (\ref{convs}) are
indeed satisfied.

To establish the first line of (\ref{convs}), we consider the consequences of
the relations (\ref{rel3}) and (\ref{rel4}).
Similar relations should also hold for other complex structures.
In terms of real quantities, (\ref{rel3}), (\ref{rel4}) and the corresponding
relations for other complex structures can be written
\begin{eqnarray} \label{con3}
&&(1+iJ^{(s)}) \nabla (1+iI) K_i^{(s)} = 0, \nn
&&(1+iJ^{(s)}) \nabla (1-iI) K_i^{(s)\prime} = 0, \qquad\mbox{(no sum over }s)
\end{eqnarray}
where $i=1,2$, $s=1,2,3$ and $I$ is the complex structure of the index bundle
introduced in (\ref{cI}). Since all quantities in (\ref{con3}) are now real,
the real and the imaginary parts should hold separately and we have the
following 12 relations,
\begin{eqnarray} \label{con4}
&&\nabla K_i^{(s)} - J^{(s)} \nabla (I K_i^{(s)}) = 0, \nn
&&\nabla K_i^{(s)\prime} + J^{(s)} \nabla (I K_i^{(s)\prime}) = 0.
          \qquad\mbox{(no sum over }s)
\end{eqnarray}
These equations are, however, not all independent.
Using the relation (\ref{ki2}), we find after some algebra that
all the relations in (\ref{con4}) can be recast into the three equations:
\begin{equation}
J^{(s)}\nabla v^a = -\nabla t_{(s)}^a,
\end{equation}
where $v = K_1 - K'_2$ as before and
\begin{eqnarray}
t_{(1)} &=& I(K'_1 - K_2), \nn
t_{(2)} &=& -K'_1 - K_2, \nn
t_{(3)} &=&  I(K_1 + K'_2).
\end{eqnarray}
This is precisely the first equation of (\ref{convs}).

With the above identification for $t_{(s)}$, the section $v^a$ turns out
to be orthogonal to $t_{(s)}^a$, i.e.,
\begin{equation} \label{vto}
v_a t_{(s)}^a = 0,
\end{equation}
which automatically satisfy the second line of (\ref{convs}).
This can be shown by using the similar arguments as in (\ref{kikj})
and the details are omitted.

As for the last two relations in (\ref{convs}), consider the identity
\begin{equation}
G^m \nabla_m K_{i\bar A}
= \frac1{\sqrt2} \int d^3x\; \Psi_{\bar A}^\dagger \gamma^0 G^m s_m
 \Ds\bar M_i \epsilon_+,
\end{equation}
which can be easily seen from (\ref{dhem}). Using the relation
$[s_m,\Ds] = \delta_mW_\mu\Gamma_\mu$, $\Ds$ can move to the left and
kills $\Psi_{\bar A}^\dagger$ since it is a zero mode. Then
\begin{eqnarray}
G^m \nabla_m K_{i\bar A}
&=& \frac1{\sqrt2} \int d^3x\; \Psi_{\bar A}^\dagger \gamma^0 G^m
\delta_mW_\mu\Gamma_\mu \bar M_i \epsilon_+ \nn
&=& -\frac1{\sqrt2} \int d^3x\; \Psi_{\bar A}^\dagger \gamma^0 D_\mu
\bar a \Gamma_\mu \bar M_i \epsilon_+,
\end{eqnarray}
where (\ref{cat}) was used in the second line. From (\ref{dhem}),
we see that the right hand side is proportional to $T_{\bar AB}$
defined in (\ref{tee}), i.e.,
\begin{equation}
G^m \nabla_m K_{i\bar A} = T_{\bar AB} K_i^B.
\end{equation}
Since the sections $v$ and $t_{(s)}$ are linear combinations of $K_i$,
this proves that the last two identities of (\ref{convs}) hold.

In summary, the low-energy effective Lagrangian (\ref{fulllag}) has all
the right properties to have $N=4$ supersymmetry.

\section{Conclusions}

In this paper, we have given a detailed derivation of the
effective action governing the low-energy dynamics of
monopoles and dyons in $N=2$ SYM
with hypermultiplets in arbitrary representations by generalizing
the techniques developed in \cite{gkly}.
This includes the case that not only adjoint Higgs fields in
the $N=2$ vector multiplet but also Higgs fields in the hypermultiplets
have non-vanishing expectation values while maintaining a
a non-trivial Coulomb branch, which is possible when the
matter representation contains a zero weight vector.
The improvement over the earlier work is that the hypermultiplets are
not necessarily in real representations. Thus we have obtained the
low-energy effective action in the most general case.
It is given by a supersymmetric quantum mechanics with potential terms which
was obtained by a non-trivial ``Scherk-Schwarz'' dimensional reduction
of (4,0) sigma models in two dimensions, which might have a more direct
stringy origin along the line of \cite{mns}.

It would be interesting to study the supersymmetric quantum mechanics derived
in this paper and check the results in the context of Seiberg-Witten
theories \cite{sw}. This has been done for example in \cite{gkpy} and
\cite{sternyi} for pure SYM case. In particular, it is an interesting 
problem to study the BPS spectrum using the effective action derived in 
this paper. 

\section*{Acknowledgments}
\noindent
We would like to thank Piljin Yi for encouragement and reading the manuscript,
and S.~Sethi for useful comments.
This work was supported by the Science Research Center Program of the
Korea Science and Engineering Foundation through the Center for
Quantum Spacetime(CQUeST) of Sogang University with grant
number R11-2005-021.

\end{document}